# Selective Laser Reaction Synthesis of SiC, Si$_3$N$_4$ and HfC/SiC Composites for Additive Manufacturing

Adam B. Peters[1], Dajie Zhang[1,2], Alberto Hernandez[1], Chuhong Wang[1], Dennis C. Nagle[1,2], Tim Mueller[1], James B. Spicer[1,2]∗

[1]Department of Materials Science and Engineering, The Johns Hopkins University
3400 North Charles Street, Baltimore, MD 21218
[2]The Johns Hopkins Applied Physics Laboratory, Research and Exploratory Development Department, 11100 Johns Hopkins Road, Laurel, MD 20723

**Abstract**

Selective laser reaction sintering techniques (SLRS) techniques were investigated for the production of near net-shape non-oxide ceramics including SiC, $Si_3N_4$, and HfC/SiC composites that might be compatible with prevailing powder bed fusion additive manufacturing processes. Reaction bonded layers of covalent ceramics were produced using in-situ reactions that occur during selective laser processing and layer formation. During SLRS, precursor materials composed of metal and/or metal oxide powders were fashioned into powder beds for conversion to non-oxide ceramic layers. Laser-processing was used to initiate simultaneous chemical conversion and local interparticle bonding of precursor particles in 100 vol% $CH_4$ or $NH_3$ gases. Several factors related to the reaction synthesis process—precursor chemistry, gas-solid and gas-liquid synthesis mechanisms, precursor vapor pressures—were investigated in relation to resulting microstructures and non-oxide yields. Results indicated that the volumetric changes which occurred during *in-situ* conversion of single component precursors negatively impacted the surface layer microstructure. To circumvent the internal stresses and cracking that accompanied the conversion of Si or Hf (that expands upon conversion) or $SiO_x$ (that contracts during conversion), optimized ratios of the precursor constituents were used to produce near isovolumetric conversion to the product phase. Phase characterization indicated that precipitation of SiC from the $Si/SiO_2$ melt formed continuous, crack-free, and dense layers of 93.7 wt% SiC that were approximately 35 μm thick, while sintered HfC/SiC composites (84.2 wt% yield) were produced from the laser-processing of $Hf/SiO_2$ in $CH_4$. By contrast, the SLRS of $Si/SiO_x$ precursor materials used to produce $Si_3N_4$ resulted in whisker formation and materials vaporization due to the high temperatures required for conversion. The results demonstrate that under appropriate processing conditions and precursor selection, the formation of near net-shape SiC and SiC composites might be achieved through single-step AM-compatible techniques.

# 1 Introduction

Silicon carbide (SiC) and silicon nitride ($Si_3N_4$) belong to the covalent family of non-oxide ceramics. This category of materials consists of nonmetallic compounds where the parent post-transition metal or metalloid atom (e.g. Si, Al, B) is covalently bonded to carbon or nitrogen through hybridization of valence electrons [1]. $Si_3N_4$ and SiC are probably the most thoroughly characterized non-oxide materials due to the breadth of structural and refractory applications [2]. By definition SiC and $Si_3N_4$, do not strictly meet the criteria for ultra-high temperature ceramics given their melting points (<3000 °C) [3]) but are significant refractories due to their unique properties. Both materials have low densities compared to transition metal non-oxides (SiC: 3.21 g/cm$^3$ and $Si_3N_4$: 3.17g/cm$^3$) and have operating temperatures up to 1200 °C or more [4]. Because of its high thermal conductivity, large bandgap (2.9 eV), high melting temperature, and excellent chemical resistance, SiC has garnered interest for utilization in high-power, high-frequency, and high-temperature devices [5]. SiC is also a common constituent of UHTC composites including ZrC-SiC, HfC-SiC, and $ZrB_2$-SiC [3], [6]. In particular, the HfC-SiC composites are considered to be potential candidates for thermal protection systems of hypersonic vehicles [7]. The addition of SiC (~20 vol%) to UHTC materials is useful in improving oxidation protection and inhibiting grain growth in hypersonic and atmospheric re-entry applications where high-temperature oxidation resistance of HfC is typically not sufficient for aerothermal flight environments [3], [8]. Similarly, $Si_3N_4$ exhibits excellent properties for structural components including the rocket nozzles on the Japanese space probe, Akatsuki, the cryogenic pump bearings of the NASA space shuttle, and recently for biocompatible orthopedic implants [9]–[11]. Table 1, summarizes the properties and applications of SiC and $Si_3N_4$.

**Table 1. Properties and Applications of Silicon Carbide and Silicon** [1], [12]–[14]

| Target SLRS Product Phase | Melting Point (Decomposition Temp) (°C) | Youngs Modulus (GPa) | Thermal Conductivity (W m$^{-1}$ K$^{-1}$) | Applications |
|---|---|---|---|---|
| SiC | 2730 | ~410-475 | 120-490 | Aerospace composites, gas turbines, structural components, power-electronics, nuclear reactors, semi-conducing materials, electromagnetic applications, abrasives [3] [15] [16] |
| $Si_3N_4$ | 1900 | ~260-310 | 25-36 | Rocket nozzles, gas turbines, machine tools, bearings, automotive turbocharges, dental and orthopedic implants [9]–[11], [17] |

Broader utilization of SiC and $Si_3N_4$ has largely been limited by the prohibitive costs and challenges in producing dense, complex components. The ability to additively manufacture robust SiC and $Si_3N_4$ for applications that require intricate geometric features would constitute a significant advancement that would reduce engineering constraints.

Compared to the many studies conducted on AM of metals, polymers, or other ceramic materials like $Al_2O_3$, those focused on SiC and $Si_3N_4$ are limited. Both SiC and $Si_3N_4$ thermally decompose at their melting temperatures of 2730 °C and 1900 °C respectively. This makes AM processing using powder bed fusion-SLM incompatible with these materials, and all well-established SiC and $Si_3N_4$ methods rely on indirect AM. These approaches typically involve SiC or $Si_3N_4$ feedstocks consolidated using organic binders during green body shaping or pyrolysis of a preceramic polymer. After binder burn out or pyrolysis, molten Si infiltration ($M_p$=1410 °C) is commonly used to increase part densities, prevent brown part disintegration, and improve mechanical properties, but this impacts the materials' refractory qualities [18], [19]. Near full densities can be obtained from multi-step techniques, but high post-processing temperatures are needed to sinter particles so parts are often subject to anisotropic consolidation, shape distortion, and significant volume changes [20].

Fewer studies have been conducted for AM of $Si_3N_4$ than for SiC. Generally, pyrolysis of pre-ceramic materials or binders are useful for carbide formation, but they are not broadly applicable to nitrides so other methods must be used. $Si_3N_4$ has been obtained through laser melting of Si, followed by post-AM nitridation to form $Si_3N_4$ with densities ranging from 63-85 vol% [21], by direct laser sintering of $Si_3N_4$ to form ultra-porous (20% dense) materials via SLS, or similar multi-step processes involving green part formation from $Si_3N_4$-binder mixtures. Table 2 summarizes different approaches to forming SiC and $Si_3N_4$ using AM and their associated challenges [22].

**Table 2. Summary of Relevant Additive Manufacturing Processes for SiC and $Si_3N_4$** (adapted from [22] and expanded using other sources).

| | Feedstock Type | AM Process | Process Type | Consolidation/Densification Process | Challenges | Citations |
|---|---|---|---|---|---|---|
| SiC | Pre-Ceramic Polymer | Stereolithography | Indirect | Polymer pyrolysis | Impurities, amorphous phases, shrinkage | [23],[24] |
| | | Direct ink writing | Indirect | | Impurities, shrinkage (>25%) during pyrolysis, dimensional restrictions; 70%+ weight loss) | [25],[26] |
| | Powder Based | Binder Jetting | Indirect | Polymer impregnation and pyrolysis; reaction bonding by molten Si infiltration; chemical vapor infiltration | Impurities, amorphous phases residual Si, multiple infiltrations | [27]–[29] |
| | | Stereolithography | Indirect | Polymer impregnation and pyrolysis | Shrinkage during pyrolysis, dimensional restrictions | [30] |
| | | Direct ink writing | Indirect | Liquid phase sintering | Shrinkage during densification | [31] |
| | | Gel-Casting with polymer skeleton | Indirect | | Shrinkage during densification (~15%), high-temp, long post-sintering required | [22] |
| | | Laminate Object Manufacturing | | | Residual Si, dimensional restrictions | [32] |
| | | Selective Laser Sintering (SLS) | Indirect | Reaction sintering of silicon and carbon + post | Silicon impurity, shrinkage during densification | [33] |
| | | Selective Laser Reaction Sintering (SLRS) | Direct | Direct reaction sintering using reactive gas | Residual Silicon, volume change with single component precursor. | [34], [35] |

**Table 2 (continued). Summary of Relevant Additive Manufacturing Processes for SiC and $Si_3N_4$** (adapted from [22]).

| | Feedstock Type | AM Process | Process Type | Consolidation/Densification Process | Technological Challenges | Citations |
|---|---|---|---|---|---|---|
| $Si_3N_4$ | Preceramic Polymer | Stereolithography | Indirect | Polymer pyrolysis | Large shrinkage and cracking during pyrolysis, impurity phases | [36], [37] |
| | Powder Based | Binder Jetting | Indirect | Debinding and reaction sintering | Low density, oxycarbides and oxynitride impurity phases | [38] |
| | | Materials Extrusion | Indirect | Addition of sintering aid and firing | Poor resolution, linear shrinkage often up to ~30% | [39], [40] |
| | | Selective Laser Sintering/Melting (SLS/SLM) | Indirect | SLS of ceramic + binder; or Si with post nitridation; direct sintering of ultra-porous materials | | [21], [41], [42] |
| | | Selective Laser Reaction Sintering (SLRS) | Direct | Direct reaction sintering using a reactive gas | | [43] |

Selective laser reaction sintering (SLRS), is investigated as a potentially alterntive processing approach for SiC and $Si_3N_4$ and other composite materials like HfC/SiC [35], [43], [44]. SLRS is based on reaction synthesis, unlike SLS/SLM which are chemically inert processes (where a feedstock material is processed in Ar or $N_2$ to create parts of the same chemical composition as the powder bed). As selective irradiation occurs, conversion to refractory non-oxides is achieved through gas-solid (or gas-liquid) nitridation or carburization reactions between heated precursor particles (metals or metal oxides) and gas-decomposition products (C from $CH_4$ or N from $NH_3$). Here, reaction bonding is thought to serve as a primary mechanism of particle adhesion under photothermal irradiation since precursor components are converted by thermally-initiated, *in-situ* chemical reactions within a reactive atmosphere [35], [43], [44].

The first studies on SLRS were conducted by Birmingham on the carburization (using $C_2H_2$) or nitridation (using $NH_3$) in Si for the formation of SiC and $Si_3N_4$ [34], [43]. These investigations utilized either $CO_2$ (25W) or Nd:Yag (150W) lasers to synthesize ≥85 wt% SiC and ~85 wt% $Si_3N_4$ in single and multi-layer coupons (~5mm x 5mm x 1mm coupons). SLRS samples were able to be handed but were porous, had minimal mechanical properties, and had features indicative of residual stresses [34], [43]. This work expands on the initial studies by Birmingham towards the formation of crack-free, non-oxide SLRS layers of SiC, $Si_3N_4$, and HfC-SiC composite materials using isovolumetric precursor formulations. Factors such as carbon and nitrogen diffusion kinetics into solid or melted precursor particles are analyzed in the context of product layer morphology and applicability to reactive additive manufacturing techniques.

## 2 Methods
### 2.1 Precursor Selection, Preparation - Benchtop SLRS

For preliminary SLRS conversion, Si and its oxide phases ($SiO_2$ or SiO) were selected based on their commercial availability and thermodynamic favorability for the for-reaction synthesis techniques. The carbonization and nitridation reactions of Si-based precursors are provided in Table 3 and Table 4. Near net-shape metal/metal oxide ratios for isovolumetric were conversion can be calculated according to Eq. 1, [21], [44], [45]:

$$f_{\Delta V=0} = \frac{v_{mo} - y v_{mx}}{v_{mo} + (1-y)v_{mx} - v_m} \qquad \text{Eq. 1}$$

where *f* is the mole fraction of metal in the composite precursor for zero volume change, $v_{mx}$ is the molar volume of the metal carbide or nitride, $v_m$ is the molar volume of the metal, $v_{mo}$ is the molar volume of the oxide, and *y* is the oxidation state of the metal species in the product (i.e +2) [21], [44], [45].

**Table 3. Isovolumetric Formation of SiC and $Si_3N_4$ using Gas-solid Reactions.**

| Product ($M_P$) | Solid Phase Precursor | Solid Precursor Mp (°C) | Reaction | Reaction with $CH_4$ | Est. Spontaneous Reaction Temp. (°C) | Volume Change (%) | M/M-O Ratio for Net-Shape Conversion in Mol. (wt%) |
|---|---|---|---|---|---|---|---|
| SiC | Si | 1410 | 1 | $Si + CH_4 \rightarrow SiC + 2H_2$ | <25 | +3.2% | 96/4 (92/8 wt%) |
| (2730°C) | $SiO_2$ | 1710 | 2 | $SiO_2 + 3CH_4 \rightarrow SiC + 2CO + 6H_2$ | 1036 | -44.9% | 96/4 (92/8 wt%) |
| $Si_3N_4$ | Si | 1410 | 3 | $3Si + 4NH_3 \rightarrow Si_3N_4 + 6H_2$ | <25 | +20.6% | 71/29 (61/39 wt%) |
| (1900°C) | SiO | 1702 (est.) | 4 | $SiO + 4NH_3 \rightarrow Si_3N_4 + 3H_2O + 3H_2$ | - (<25°C for $SiO_{(g)}$) | -29.4% (est) | 71/29 (61/39 wt%) |

**Table 4. Isovolumetric Formation of HfC/SiC Composite Gas-solid Reactions.**

| Product ($M_P$) | Solid Phase Precursor | Solid Precursor Mp (°C) | Reaction | Reaction with $CH_4$ | Est. Spontaneous Reaction Temp. (°C) | Volume Change (%) | M/M-O Ratio for Net-Shape Conversion in Mol. (wt%) |
|---|---|---|---|---|---|---|---|
| HfC/SiC Composite | Hf | 2227 | 1 | $Hf + CH_4 \rightarrow \mathbf{HfC} + 2H_2$ | <25 | +14.3% | 84/16 (94/6 wt%) |
| (2730°C) | $SiO_2$ | 1710 | 2 | $SiO_2 + 3CH_4 \rightarrow SiC + 2CO + 6H_2$ | 1036 | -44.9% | 84/16 (94/6 wt%) |

Thermodynamic calculations and preliminary experiments suggested that Si and or $SiO_2$ might be viable for SLRS synthesis of SiC. However, the high spontaneous reaction temperatures for $Si_3N_4$ synthesis from $SiO_2$ in $NH_3$ are not favorable. The high minimum conversion temperature (1924 °C) for $Si_3N_4$ formation and low $SiO_2$ precursor absorptivity at the laser wavelength (445 nm) prevented conversion of $SiO_2$ in preliminary tests. Rather, silicon monoxide (SiO) was selected as an alternative precursor to $SiO_2$ for $Si_3N_4$ production. Previous investigations indicated that particle size reduction is favorable for reactivity [45]. Commercially available precursors Si, SiO, and $SiO_2$ (all <44 μm, -325 mesh) were selected. SEM imaging in Fig. 1 indicates average particle sizes were ~5-25 μm which could facilitate conversion and minimize consolidation when compared to the sub-micron particles. The SiO precursor material was sieved to remove particles larger than 44 μm. All two-component precursor systems were mixed in a roller mixing machine for 2 hrs to ensure homogeneous distribution particles. Particle morphologies and the optical characteristics of each precursor material are summarized in Table 5.
.

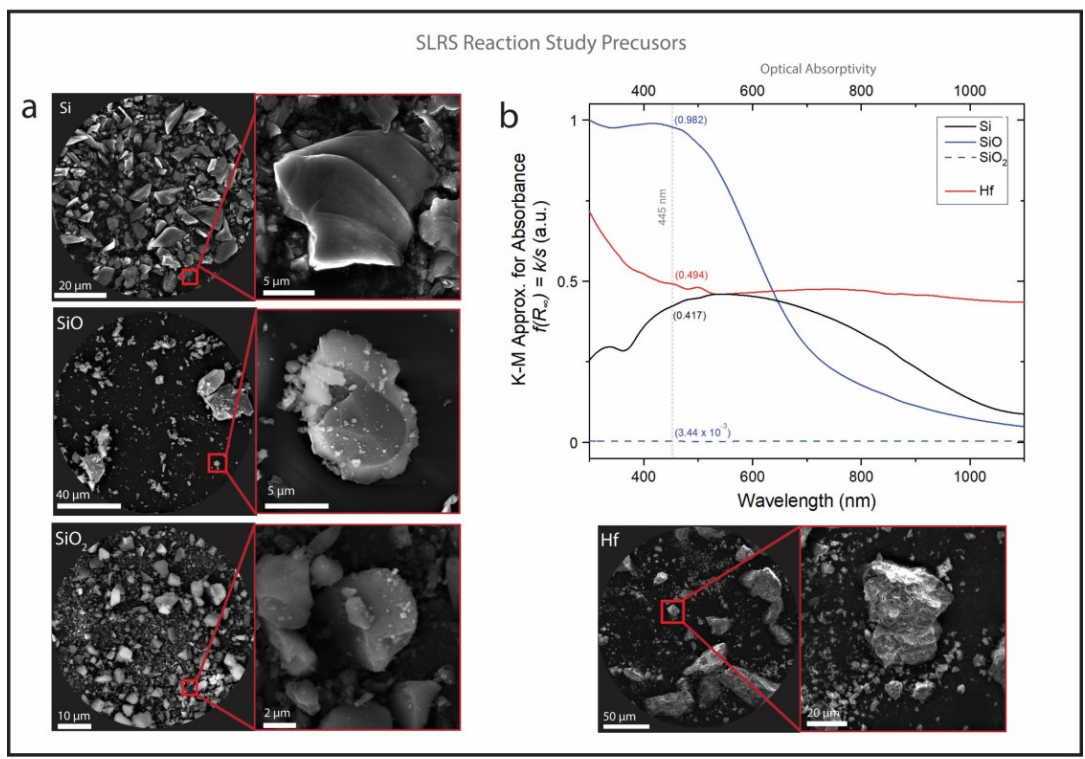

**Fig. 1.** (a) SEM images of silicon and silicon oxide precursor materials used for SLRS processing. The lower magnification image on the right illustrates the variation in particle size for a given precursor, while the image on the left shows typical particle morphology. (b) The optical absorbance of metal and metal oxide precursors from 300-1000 nm was measured using diffuse reflectance UV-vis spectroscopy. Here the diffuse reflectance is presented using normalized Kubelka-Munk absorbance spectra (k/s), which is the ratio of absorbance and scattering coefficients k/s. The spectra were normalized to the highest K-M value between 300-1100 nm.

**Table 5. Selected Precursor Powder Characteristics for Reactivity Studies**

| Chemistry | Feedstock Powder | Chemical Purity (wt. %) | Manf. Mesh Size | Packing Density | Particle Morphology |
|---|---|---|---|---|---|
| Si | Sigma Aldrich | 99% | -325 (<44 μm) | 0.42 | Jagged, Sharp, Dense |
| SiO | NOAH Technologies | 99.5% | -325 (<44 μm) | 0.63 | Rough, Dense |
| $SiO_2$ | Sigma Aldrich | 99.5% | -325 (<44 μm) | 0.24 | Rough, Dense |
| Hf | Atl. Equip. Engineers. | 99.5% | 0.50 | Rough, Blocky, Dense | Rough, Blocky, Dense |

### 2.1.1 *Experimental Process - Preliminary Studies Using SLRS*

Syntheses were conducted using various silicon-based precursor materials to understand the reaction conditions that facilitate SiC and $Si_3N_4$ formation. Silicon and silicon oxide-containing powders were poured into cylindrical reservoirs (stainless steel, inner diameter: 25.5 mm, depth: 4.5 mm) and screened using a metal blade each precursor bed into an even layer, creating a flush powder surface representative of powder bed fusion. Each precursor bed was then inserted into the laboratory scale SLRS reaction chamber shown in Fig. 2 for selective laser processing.

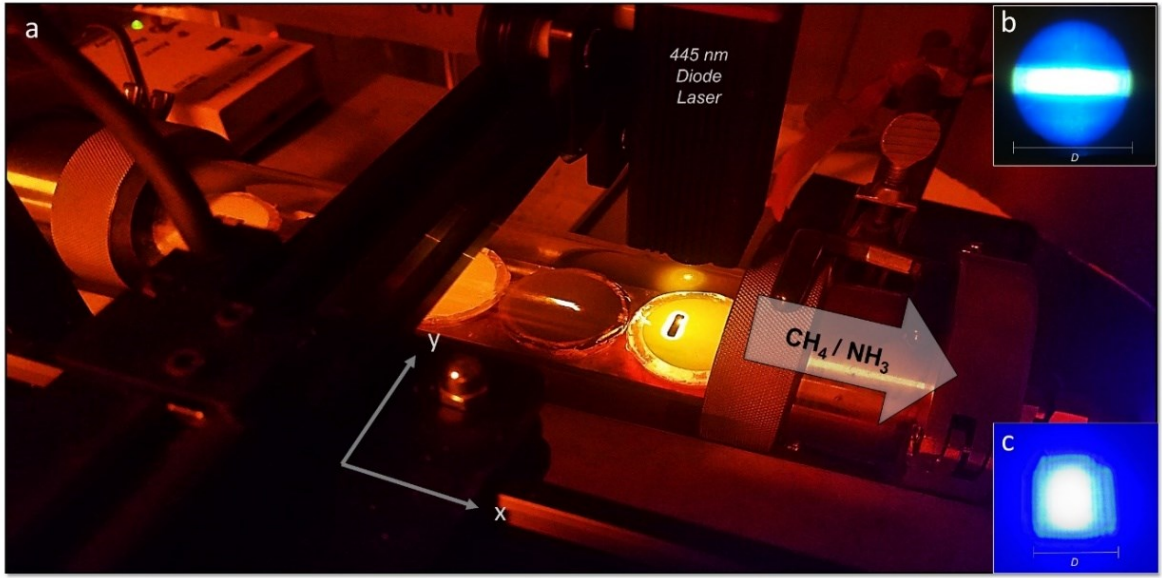

**Fig. 2.** (a) Photograph of the SLRS apparatus. Metal and metal oxide powders are in reservoirs in the quartz tube. (b) The beam profile of the 4 W, 445 nm diode laser used for conversion of single component precursor materials. (c) The beam profile of the 5.25 W, 445 nm diode laser was used for the conversion of various precursor materials using the "high power" setting.

Lightburn software was used to control the bi-directional scan path of the laser beam on an X-Y gantry system to produce 12 mm x 12 mm scanned regions on the powder bed surface. The low power (3 W) processing was used as a baseline to assess the conversion propensity of single-component materials by gas-solid reactivity. Laser power was increased to investigate precursor melting and gas-liquid conversion mechanisms and their applicability to additive manufacturing. A complete summary of the laser processing conditions is provided in Table 6.

**Table 6. SLRS Laser Processing Parameters.**

|  | **Low Power** | **Medium Power** | **High Power** |
|---|---|---|---|
| Materials | Si, SiO, $SiO_2$ | $Si/SiO_2$; Si/SiO | Si/SiO |
| Processing Gas | 100 vol.% $NH_3$/$CH_4$ | 100 vol.% $NH_3$/$CH_4$ | 100 vol.% $NH_3$ |
| Products | SiC, $Si_3N_4$ | SiC, $Si_3N_4$ | $Si_3N_4$, HfC-SiC |
| Wavelength | ($\lambda$) = 445nm | ($\lambda$)= 445nm | ($\lambda$)= 445 nm |
| Average Power | (P) = 3.0 W | (P) = 4.0 W | (P) = 5.25 W |
| Spot Diameter. | (D) = ~400 μm | (D) = ~750 μm | (D) = ~300 μm |
| Energy Density* | (E) = 8W/$mm^2$ | (E) = 32W/$mm^2$ | (E) = 58W/$mm^2$ |
| Processing Speed | (V) =100 mm/min | (V) = 100 mm/min | (V) = 100 mm/min |
| Gas Flow rate | 250 $cm^3$/min | 250 $cm^3$/min | 500 $cm^3$/min |

*Note, some transient optical output fluctuations (+/- 10%) were observed during processing.

A translational scan speed of 100 mm/min was used for all trials. After precursor materials were placed in the reaction chamber, the vessel was purged with high-purity argon (500 SCCM) to prevent oxidation during laser heating. The reaction chamber was then reconstituted with flowing 100 vol% $CH_4$ (250 SCCM) for SiC, SiC/HfC or 100 vol% $NH_3$ (500 SCCM) for $Si_3N_4$ synthesis. Flow rates of 500 SCCM were used for reactions where the laser power exceeded 4 W which initiated materials vaporization. For each reaction, square 12mm x 12mm single layer coupons of SiC or $Si_3N_4$-containing were produced for chemical and microstructural characterization. Precursor materials processed in Ar (control), $CH_4$ or $NH_3$ were characterized using X-ray diffraction and Rietveld refinement. XRD characterization was performed on processed regions while supported by their powder beds due to the fragility of the samples. $SiO_2$ and SiO precursor materials do not produce defined crystalline peaks in the XRD diffractogram between 20-120° 2θ so phase data are summarized with respect to total crystalline content. Quantitative phase characterization of crystalline phases was performed from 20° to 80° 2θ using a Rietveld refinement with Match! software. The weighted profile R-factor ($R_{wp}$) and $\chi^2$ for each sample were minimized to ensure the quality of fit for quantitative phase characterization. SEM microscopy was conducted using a Tescan Mira 3 GM Scanning Electron Microscope (Kohoutovice, 62300, Czech Republic) on sample surfaces and on epoxy-mounted cross-sections using procedures described previously [21].

## 2.2 Results and Discussion

To investigate selective laser reaction sintering of covalent non-oxides, Si and $SiO_x$ precursors were processed separately in the presence of 100 vol.% $CH_4$ or $NH_3$ gas. Results from irradiation of Si, SiO and $SiO_2$ are presented in Table 7. Photographs of the selectively processed 12 mm x 12 mm powder bed areas are shown in Fig. 3.

**Table 7. X-Ray Composition Analysis Using Rietveld Refinement Modeling: Conversion of Si or $SiO_x$ Precursor Particles in $CH_4$ or $NH_3$**

| Target Product | Solid Phase Precursor | Stoichiometric Reaction with $CH_4$ | Est. Spontaneous Reaction Temp. (°C) | Total Carbide or Nitride (wt%) | SiC or $Si_3N_4$ Product Phase | Unr. Si (wt%) | Carbon Residue |
|---|---|---|---|---|---|---|---|
| SiC | Si | $Si + CH_4 \rightarrow SiC + 2H_2$ | <25 | ~82.6 | SiC (3C): 70.2% <br> SiC (96R): 12.4 | 17.4% | - |
| | SiO | $SiO + 2CH_4 \rightarrow SiC + CO + 4H_2$ | - | 33.5%* | SiC (3C): 26.8% <br> SiC (96R/6H): 6.7% | 66.5% | Yes |
| | $SiO_2$ | $SiO_2 + 3CH_4 \rightarrow SiC + 2CO + 6H_2$ | 1036 | 100%* | SiC (3C): 86.2% <br> SiC (96R/6H): 13.8% | 0% | Yes |
| $Si_3N_4$ | Si | $Si + 4NH_3 \rightarrow Si_3N_4 + 6H_2$ | <25 | 97.7% | $Si_3N_4$ (α): 72.4% <br> $Si_3N_4$ (β): 25.3% | 2.3% | - |
| | SiO | $SiO + 4NH_3 \rightarrow Si_3N_4 + 3H_2O + 3H_2$ | - | 47.6%* | $Si_3N_4$ (α): 30.2% <br> $Si_3N_4$ (β): 17.4% | 52.4% | - |
| | $SiO_2$ | $3SiO_2 + 4NH_3 \rightarrow Si_3N_4 + 6H_2O$ | 1924 | 0 % | - | 0% | - |

*Calculated based on total crystalline materials quantified via Rietveld refinement and does not include non-crystalline phases. Non-oxide may contain trace oxygen contaminants in the form of oxycarbide species [46].

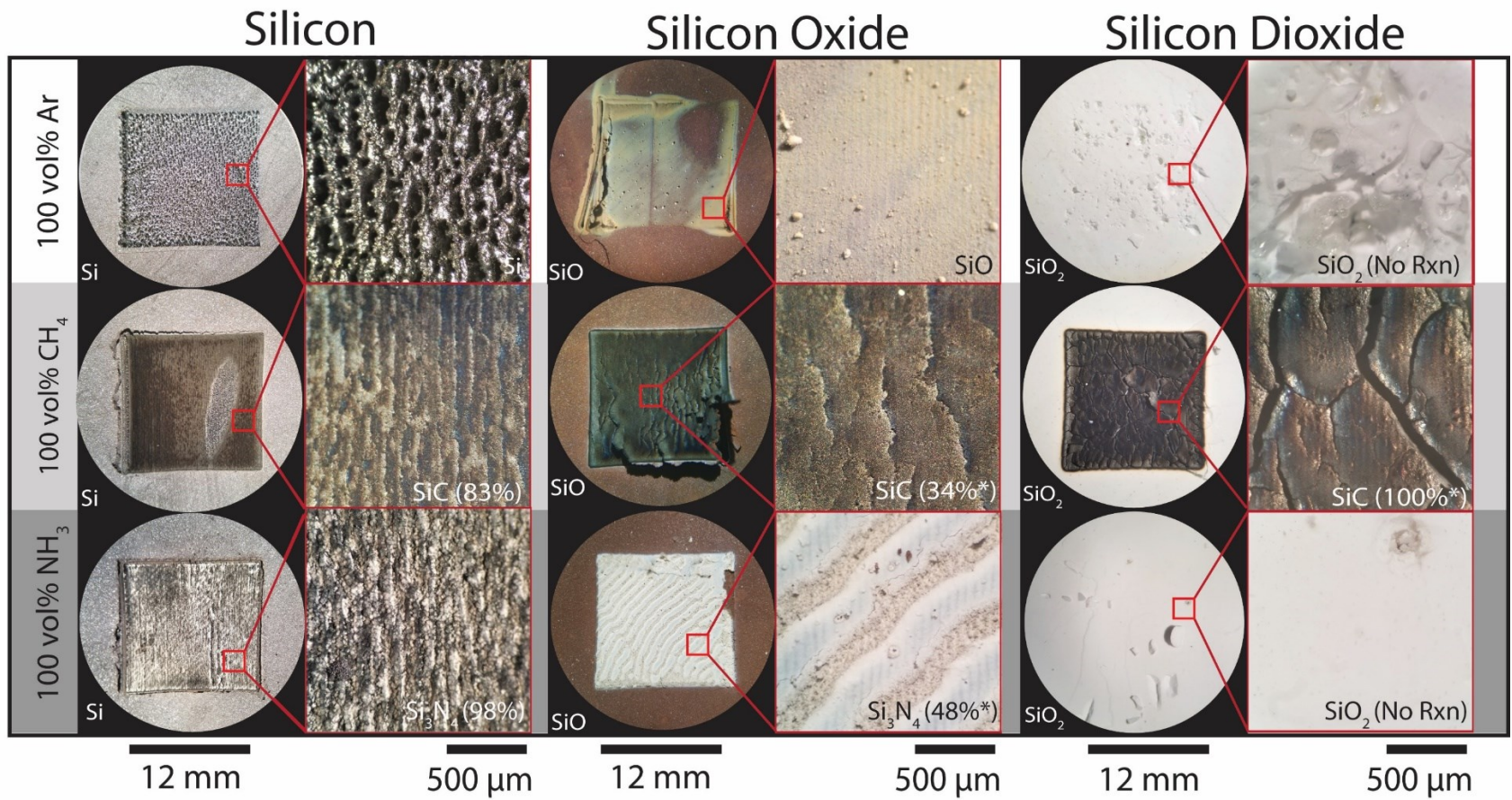

**Fig. 3.** Photomicrographs of silicon or silicon oxide precursor powders irradiated by the 3.0 W laser at 100% power (100 mm/min scan speed) in 100 vol.% Ar (control), $CH_4$, or $NH_3$. The <u>target</u> non-oxide products are given in the lower right corner of each image and indicate the total wt% of crystalline carbides or nitrides formed. In the cases of SiO and $SiO_2$, non-crystalline phases were not able to be accurately quantified, the relative weight percentage of crystalline material is indicated on each micrograph. The ellipsoidal region in the center of the Si sample processed in $CH_4$ is related to non-uniform laser irradiation of the powder bed.

### 2.2.1 *Selective Laser reaction Synthesis of SiC*

*2.2.1.1 Laser processing of Si in CH$_4$*

Silicon and silicon oxides can readily be converted to carbides using reactive laser processing. Others have indicated the reactive laser synthesis of SiC by reaction of CH$_4$ from either powders [34] or bulk single crystals [5], however, reaction synthesis techniques have not been developed using the incorporation of other precursor materials that might mitigate volume changes for AM applications. Broadly, the combined results in Table 7 and XRD in Fig. 5 indicate that SLRS processing of Si produced significant conversion of precursor materials to SiC. The phase results reveal the formation of approximately ~82.6 wt % SiC through the formation of different allotropes: 3C-SiC (70.2 wt%) and 96R/6H-SiC (12.4%). The combined results are consistent with the initial study by Birmingham, where up to 80-85 wt% SiC was produced from the irradiation of <44 μm powder using two passes with 2.3 W of laser energy (avg.) [34]. While no differentiation between the 3C and secondary SiC phase(s) was noted, both Fig. 5 and results presented by Cheung [34] contain a primary SiC peak at ~35.7° 2θ and left shoulder at 33.6 ° 2θ (1-0-0 peak) are indicative of 96R or 6H-SiC and/or similar SiC polymorphs with hexagonal or trigonal structure [47], [48]. Due to the 250+ polymorphs of SiC, it was challenging to unambiguously confirm the specific secondary phase [49].

The high optical absorptivity of Si at 445 nm, favored efficient photothermal energy conversion, high surface temperatures, and selective melting of silicon (M$_p$= 1410 °C) when processed in Ar. Photomicrographs in Fig. 3 of Si-SLS (in Ar) and Si-SLRS (in CH$_4$) show strikingly different surface morphologies that provide insight into their respective consolidation mechanisms. SEM images in Fig. 4 reveal that Si particles reacted with CH$_4$, were well bonded, and likely formed by the partial-melting, solution-precipitation mechanism as described by Cheung [34]. Moreover, the needle-like formations could be the result of heterogeneous SiC nucleation from vaporized Si (3265 °C) that were rapidly heated as a consequence of Si's high optical absorptivity and moderate vapor pressures [50]. These formations should be compared to the typical melting behavior observed for Si-Ar processing.

For a Si-CH$_4$ gas-liquid reaction, the concentration profile of carbon into the Si melt pool might be determined by non-steady state diffusion,

$$C_m(x,t) = C_o(1 - \text{erf}\left(\frac{y}{\sqrt{4Dt}}\right)) \qquad \text{Eq. 2}$$

where *y* is the distance into the melt, $C_o$ is carbon concentration at $t = 0$, $C_m$ is the concentration at time *t*, and *D* is the diffusivity of carbon in molten Si (5 x 10$^{-4}$ cm$^2$/s [51], [52]) [35], [53], [54].

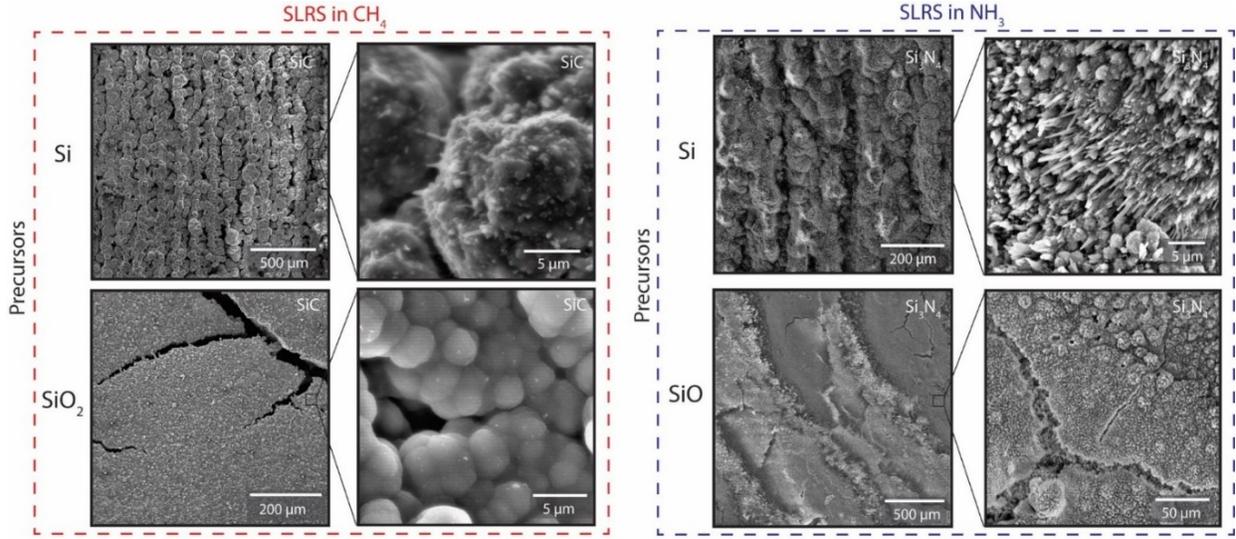

**Fig. 4.** SEM micrographs showing the surface microstructure of SLRS processed Si, SiO, SiO$_2$ in CH$_4$ or NH$_3$. Target product phases are labeled in the upper right corner of each SEM image.

Using an assumption of rapid, near-instantaneous solidification of molten Si following the laser scan, the laser interaction zone ($d_{int}$ ~ 400 μm) might remain liquid for ~0.24 seconds following laser heating in the presence of CH$_4$. Likely, the rate of CH$_4$ decomposition is not the rate-limiting factor; according to Eq. 2, 50 at.% saturation would occur at 107 μm into liquid Si. This diffusion distance exceeds the diameter of the Si precursor particles (<44 μm) suggesting that carbon could saturate the melt pool with adequate laser penetration into the powder bed. This result generally seems to agree with experimental findings from phase quantification where ~83% SiC was produced within the characterization volume of the XRD (x-ray penetration depth for $I/I_{o\text{-}SiC}$ = 0.05 is 200 μm).

Microstructural assessment of SLRS processed-Si suggests that the accumulation of stresses was minimal compared to the conversion of metal precursors. The maximum volume change for conversion of Si to SiC is +3.4% and might be accommodated by residual porosity in the powder bed. While single layers of SLRS of SiC-containing materials were generally crack-free, accumulation of tensile or thermal stresses during multi-layer fabrication in initial SLRS studies [55] may have caused the interlay delamination and poor mechanical properties reported.

*2.2.1.2 Laser processing of Silicon Oxides in CH$_4$*

Synthesis of SiC from SiO$_2$ requires progressive reduction through a SiO intermediate according to Eqs. 3-4 [56].

$$\text{SiO}_{2(s)} + \text{CH}_{4(g)} \rightarrow \text{SiO}_{(g)} + \text{CO}_{(g)} + 2\text{H}_{2(g)} \quad \text{Eq. 3}$$
$$\Delta G_r = 757.07 - 0.4373T \text{ (kJ)}$$

$$\text{SiO}_{(g)} + 2\text{CH}_{4(g)} \rightarrow \text{SiC}_{(s)} + \text{CO}_{(g)} + 4\text{H}_{2(g)} \quad \text{Eq. 4}$$
$$\Delta G_r = 99.11 - 0.2160T \text{ (kJ)}$$

During SLRS, CH$_4$ dissociation provides the carbon source for SiC formation from silicon oxides. Schei and Halvorsen note that if the molar ratio of available carbon to SiO$_2$ is less than 3, silicon can be lost through low vapor pressure SiO$_{(g)}$. SiC can then be formed from the interaction of SiO$_{(g)}$ and dissociated carbon C [57].

$$SiO_{2(s)} + xCH_{4(g)} \rightarrow \frac{(x-1)}{2} SiC_{(s)} + \frac{(3-x)}{2} SiO_{(g)} + \frac{(1+x)}{2} CO_{(g)} + 2xH_{2(g)} \qquad \text{Eq. 5}$$

Minimal vaporization of SiO$_{2-x}$ into the gas stream or onto the quartz tube was observed. This qualitative observation indicates that the laser-assisted reaction might be rapid and carbon readily available to prevent the loss of material through the formation of SiO$_{(g)}$. XRD phase characterization for the SLRS processing of SiO and SiO$_2$ in CH$_4$ suggests that significant phase fractions SiC can be produced from either precursor. For conversion of SiO$_{(s)}$, a product yield ratio of 34/66 wt% SiC/Si was obtained. Meanwhile, for the reduction of SiO$_{2\,(s)}$, SiC was the only crystalline material found in the diffractogram. In both cases, the lack of diffraction peaks for SiO and SiO$_2$ made the precise quantification of oxides difficult. The presence of the Si in Fig. 5 for SLRS processing of SiO in CH$_4$ (and NH$_3$), suggests that the reaction in Eq. 5.3 occurs concurrently with the thermally catalyzed disproportionation of Si:

$$2SiO(s) \rightarrow 2SiO_{(g)} \rightarrow SiO_{2(s)} + Si_{(s)} \qquad \text{Eq. 6}$$

Depending on the local temperature, disproportionation products can react to form SiC through Eq. 3 and 4 (this reaction occurs at experimental temperatures between ~1400 - 1975 °C) or by Si carburization for the Si-O-C system [56]. Eq. 6 may contribute to higher SiC yield, where the formation of solid SiO$_2$ and Si could prevent materials loss into the gas stream. The microstructures in SEM micrographs of laser processed SiO$_2$ in Fig. 4 support this process. Additionally, the increased x-ray counts for 3C-SiC peak (3023 counts) at 35.67° 2θ for SLRS of SiO$_2$ vs. SiO (949 counts) suggests that greater SiC yield by mass despite lower optical absorptivity.

The use of 100 vol% CH$_4$ may also favor conversion by providing excess C to favor reactivity thereby suppressing SiO vaporization (Eq. 5) and whisker formation that has been observed by others [58], [59]. This excess carbon availability is apparent in XRD results where carbon or graphite is indicated by the broad peak at approximately 26.6° [60]. The increased peak intensity at this location in Fig. 5 might also be attributed to trace silicon oxycarbide species that produce reflections from the (0-0-2) planes of graphite [60].

Compared to traditional carbothermal reduction techniques of SiO$_2$ and C$_{(s)}$, laser processing of SiO$_2$ proved to be more efficient than traditional furnace processing. Li found that carbothermal reduction of SiO$_2$/C mixtures in low CH$_4$-concentration atmosphere was slow: ~75% SiC yield was produced after 150 mins at 1550 °C with a significant fraction of Si lost to SiO vaporization [59]. Silicon loss through SiO vaporization became negligible when CH$_4$ gas concentrations were increased to ~50/50 vol% CH$_4$/Ar [59]. Higher temperatures accelerate the rate of reduction. Even though the exact silicon loss to vaporization from SiO$_2$ reduction was not determined in this work, SLRS processing appears to be a viable means for spatially selective SiC production in sub-second time scales. Results for SiO$_2$ reduction, shown in Fig. 4, indicate a SiC microstructure containing dense regions of reaction sintered particles with cracking related to densification and the theoretical - 44.9% volume reduction.

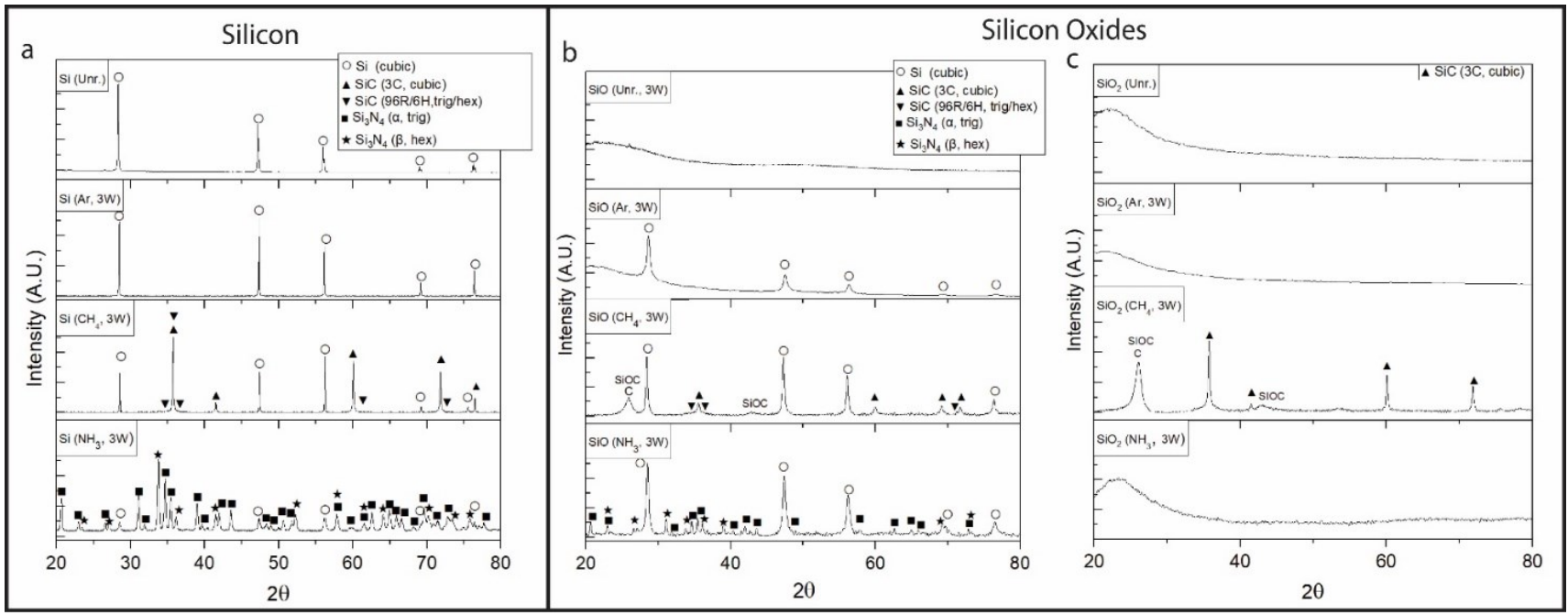

**Fig. 5**. XRD spectra of Si, SiO, and SiO$_2$ precursor materials thtat were processed using selective laser reaction sintering in 100% Ar, CH$_4$, and NH$_3$ using the 445 nm diode laser.

### 2.2.2 Selective Laser Reaction Synthesis of $Si_3N_4$

*2.2.2.1 Laser processing of Si in $NH_3$*

The selective laser reaction in 100 vol% $NH_3$ of Si precursors was performed for $Si_3N_4$ production. Typical nitridation of Si powders occurs efficiently between 1100-1500 °C, where a combination of both α- and β are almost always formed [61], [62], [63]. The reaction is strongly exothermic and must be controlled [61]. SLRS processing yielded nearly complete conversion to $Si_3N_4$ (97.7wt%) with both α-$Si_3N_4$ and β-$Si_3N_4$ having been produced using the lowest power 3 W laser setting. The yield of α-$Si_3N_4$ and β-$Si_3N_4$ were determined to be 72.4 wt% and 25.3 wt% respectively. Literature suggests that Si-N complexes typically adopt the α-$Si_3N_4$ configuration when synthesized by vapor phase reactions and or/with molecular nitrogen; β-$Si_3N_4$ is thought to form by liquid-phase reactions and or the complexing of silicon with active (possibly atomic) nitrogen [62], [64]. According to Kaiser and Thurmond [64] and Jennings [62], nitrogen does not readily dissolve in solid silicon, so direct solid-state reactions might not occur during SLRS. Rather, as the local temperature is raised, nitrogen dissociates and interacts with melted or vaporized Si molecules which are transported to the $Si_3N_4$ reacting interface. The local microstructures are shown by SEM micrographs in Fig. 4. The prevalence of α-$Si_3N_4$ agrees with findings by others where needle-like structures form and are indicative of two-step nitride synthesis: nitrogen undergoes direct adsorption followed by surface transport and/or evaporation and condensation of Si onto the tip of a growing $Si_3N_4$ needle [62], [63]. With a gradient in temperature or concentration, Si adsorption onto $Si_3N_4$ sites becomes increasingly likely with distance, further growing the formation. The vapor pressures of Si, SiO, and $SiO_2$ precursors compared to other transition metal and metal oxide precursors are given in Fig. 6. Of all materials processed in this work Si, SiO, and $TiO_2$ have the lowest temperature-dependent Antoine vapor pressure curves. They were the only materials to form whiskers through SLRS synthesis and this is shown in Fig. 4 [65]. The formation of whiskers is significant to AM as it might inhibit consistent materials deposition and formation of dense ceramic layers. Larger particle sizes can be used to suppress vaporization, but materials properties should still be considered when engineering near net-shape precursors for SLRS.

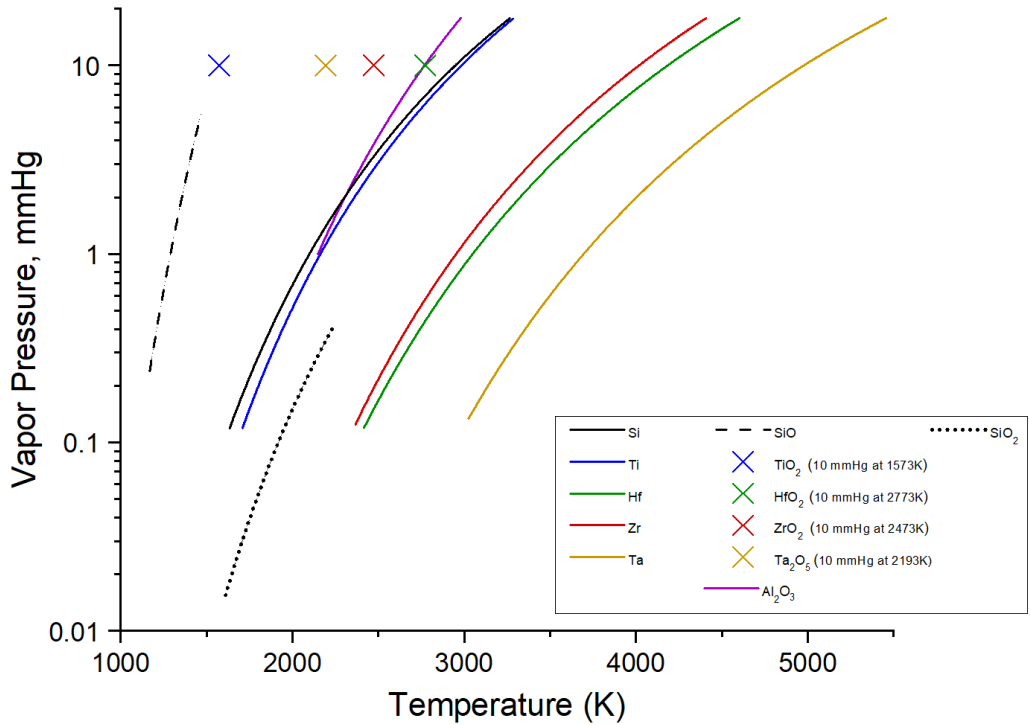

**Fig. 6.** Vapor pressures of Si, SiO, SiO$_2$, and other metal or metal oxide precursor materials that were processed by SLRS [65]–[67].

Atkinson [68] and Jennings [62], [63] noted that the overall dimensions of a powder compact do not significantly change during the nitridation of Si to reaction bonded Si$_3$N$_4$ when processed using traditional furnace heating. Over long reaction times, vapor phase Si transport fills in residual particle porosity during conversion accommodating the ~21% volume expansion [62]. This behavior contrasts with results presented in this work on laser-assisted nitride formation of interstitial non-oxides [45] where non-equilibrium, laser sintering/partial melting of low vapor pressure materials induced volume changes that led to macroscopic cracking and AM-layer defects.

As the Si$_3$N$_4$ reaction front progresses, the Si surface must be exposed to NH$_3$ for adsorption and gas decomposition. Nitrogen has no solubility in solid Si and diffusion of nitrogen through bulk α-Si$_3$N$_4$ and β-Si$_3$N$_4$ is extremely slow (with diffusion coefficients of $1.2 \times 10^{-12}$ cm$^2$/s and $6.8 \times 10^{-6}$ to $1 \times 10^{-10}$ cm$^2$/s respectively [62]). This indicates that transport of N through a Si$_3$N$_4$ product layer might not occur on timescales associated with AM so melting or vaporization is required [69]. Using Eq. 5.1, the estimated distance for N diffusion into liquid Si was determined to be only 0.85 μm during the 0.24 s irradiation time ($D_{\text{N in Si }(liq)}$: $3.2 \times 10^{-8}$ cm$^2$/s). Calculations indicate that SiC may form by gas-liquid conversion more readily than Si$_3$N$_4$ due to enhanced C diffusivity in liquid Si (~4 orders of magnitude larger, C in Si$_{liq}$: 2-$5 \times 10^{-4}$ cm$^2$/s). Comparatively, the Si migration to a Si$_3$N$_4$ reaction interface (that leads to whisker formation, $D_{\text{Si-Si}} = 9 \times 10^{-3}$ cm$^2$/s) would be preferred to depth-dependent, shrinking core conversion where N must penetrate a passivation layer. The lack of macroscopic volumetric expansion observed during SLRS processing of Si to 97.7 wt% Si$_3$N$_4$ appears to agree with a theory proposed by Atkinson and Jennings where non-solid, Si-transport and gas-phase reactivity might fill in residual powder bed porosity [62], [63], [68].

*2.2.2.2 Laser processing of SiO in NH₃*

Conversion of $SiO_2$ to $Si_3N_4$ in $NH_3$ (spontaneous at 1924 °C) was not achieved using the experimental processing conditions. Rather, selective laser processing of SiO in $NH_3$ using the 3 W laser processing conditions produced 47.6% wt% $Si_3N_4$ (30.2 wt% α-$Si_3N_4$, and 17.4 wt% β-$Si_3N_4$ β) and 52.4 wt% Si (of the characterizable crystalline phases). XRD indicates irradiation reduces SiO to Si, where $Si_3N_4$ can be formed by vapor phase reaction or through nitridation of reduced Si intermediates. Fig. 3 shows the surface microstructure of the laser processed sample where substantial SiO vaporization occurred. During the multi-step reduction process, the low vapor pressure of SiO caused vaporized products to fill the gas stream during lasing, attenuating the power of the beam. No appreciable SLRS-layer integrity or inter-particle reaction bonding was observed. Although selective laser heating of $SiO_{(s)}$ in nitrogenous-containing gases might encourage the formation of $Si_3N_4$, SLRS of mixtures containing other additives (e.g. carbon to form CO reduction by-products) might be a more effective strategy for *in-situ* nitride formation [50], [62], [63].

### 2.2.3 *Conversion of Near-Net Shape Silicon/Silicon-Oxide Precursor Materials*

Results from sections 2.2.1 and 2.2.2 show SLRS conversion of Si and $SiO_2$ by $CH_4$, and Si by $NH_3$ produced high non-oxide yields. Near net-shape precursor, mixtures were formulated in an attempt to improve the surface microstructure of SLRS samples by compensating for conversion-induced volume changes and better understand the impact of precursor-dependent gas-solid/liquid and vapor phase reactions when conditions favor complete reactivity. Laser power for processing of the precursor mixtures was increased to enhance chemical conversion by increasing temperatures at the laser-interaction zone [45].

The surface morphology processed regions are shown in Fig. 8 where optical photomicroscopy and SEM images are shown. XRD phase composition results for each sample are presented in Fig. 7 and Table 8. In addition, cross-sectional SEM micrographs for $Si/SiO_x$ mixtures are presented in Fig. 8 to inform the discussion.

**Table 8. X-Ray Composition Analysis Using Rietveld Refinement Modeling: Conversion Of Mixed Si /SiOₓ Precursor Particles in CH₄ And NH₃**

| Target Product | Solid Phase Precursor (wt%) | Laser Power (W) | Total Carbide or Nitride (wt%) | SiC or Si₃N₄ Product Phase | Unr. Si (wt%) | Notes |
|---|---|---|---|---|---|---|
| SiC | Si/SiO₂ (92/8 wt%) | 4 | 93.7% | SiC (3C): 54.6% SiC (96R): 39.0% | 6.3% | Carbon residue on the quartz tube |
| Si₃N₄ | Si/SiO (61/39 wt%) | 4 | 92.8% | Si₃N₄ (α): 71.5% Si₃N₄ (β): 21.4% | 7.2% | Mild vaporization, residue on the quartz tube |
|  | Si/SiO (61/39 wt%) | 5.25 | 78.9% | Si₃N₄ (α): 55.4% Si₃N₄ (β): 23.5% | 21.1% | Significant vaporization, residue on the quartz tube |

*Calculated based on total crystalline materials quantified via Rietveld refinement and does not include non-crystalline. Non-oxide may contain trace oxygen contaminants in the form of oxycarbides or oxynitride species [46].

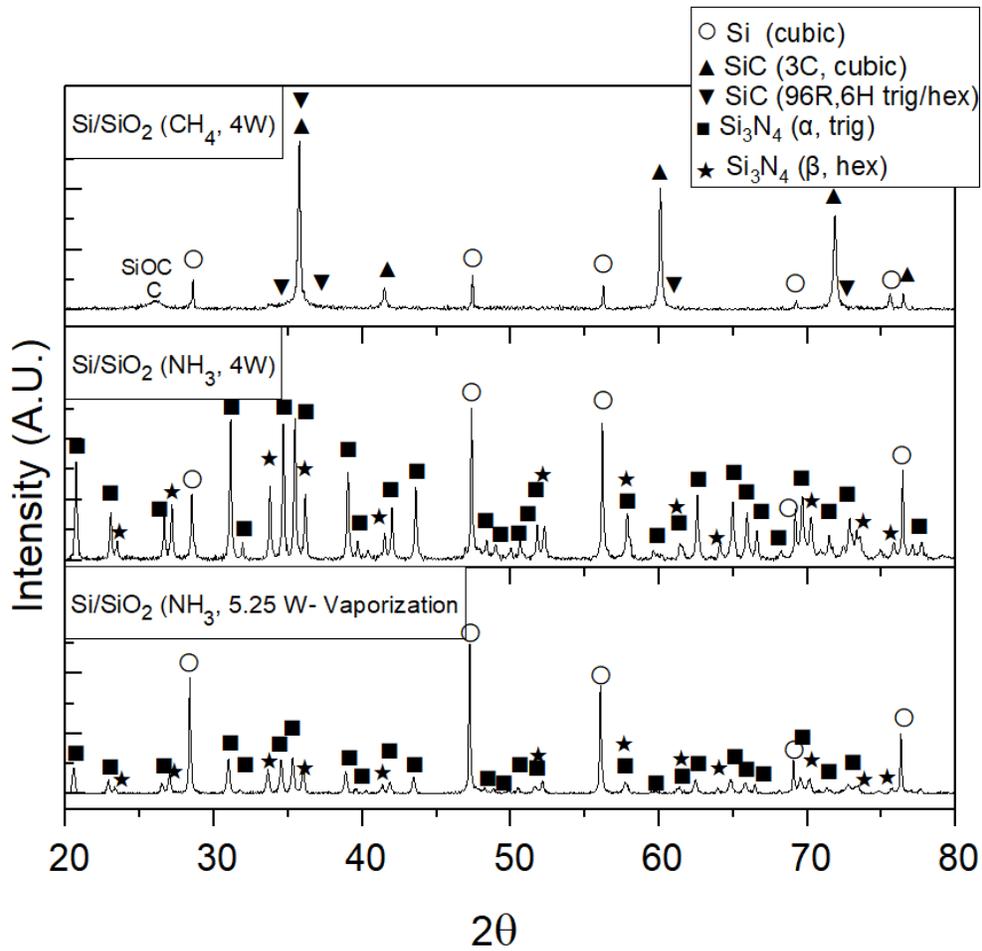

**Fig. 7**. XRD spectra of Si/SiO$_x$ precursor formulations that were processed using selective laser reaction sintering in 100% Ar, CH$_4$, and NH$_3$ using the 445 nm diode laser.

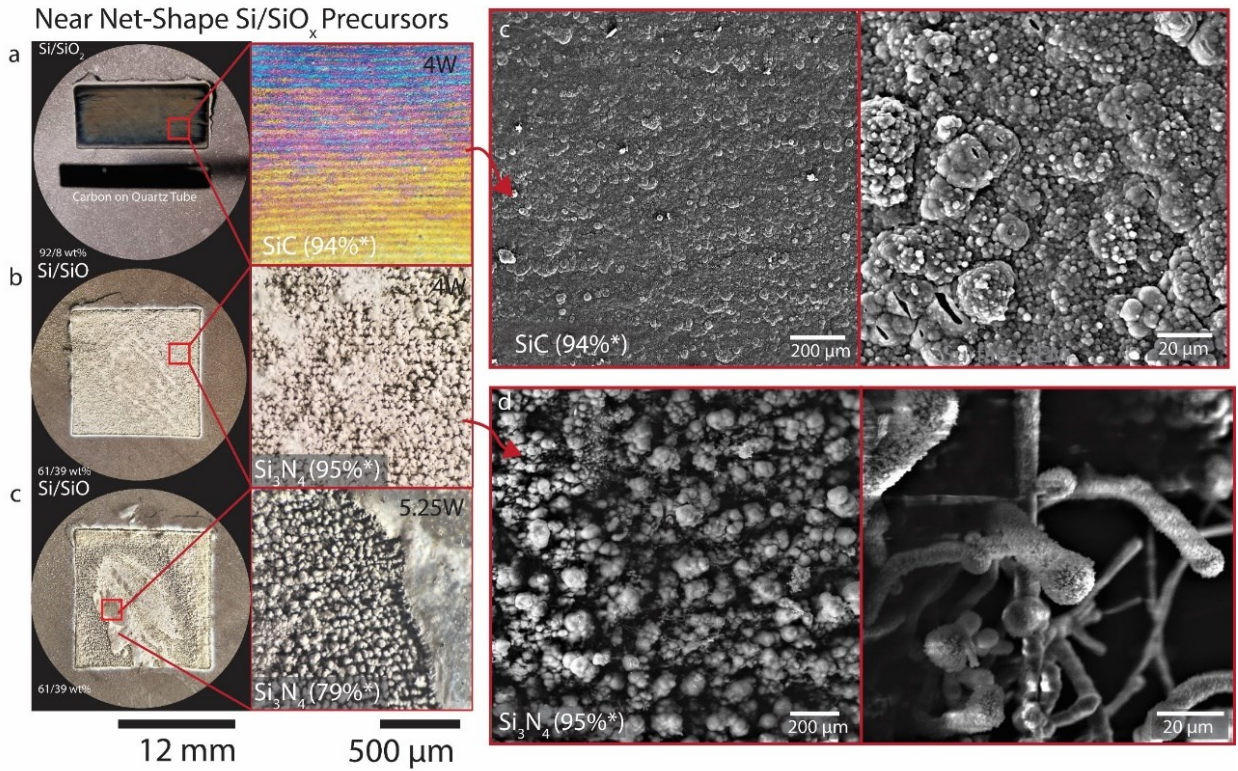

**Fig. 8.** (a-c) Photomicrographs and of Si/SiO$_x$ precursor laser sintered in CH$_4$ or NH$_3$ for SiC or Si$_3$N$_4$ formation respectively. (c-d) SEM images of showing the surface microstructure of Si/SiO$_2$ and Si/SiO precursor materials laser processed to SiC and Si$_3$N$_4$. The percentages in the lower corner of each image indicate the total wt% of SiC or Si$_3$N$_4$ of the total crystalline material.

*2.2.3.1 Laser processing of near-net-shape Si/SiO$_2$ in CH$_4$*

Figs. 7-10 show that SLRS processing of 92/8 wt% Si/SiO$_2$ in CH$_4$ (4 W) produced ~94 wt% SiC with a continuous surface and cross-sectional microstructure with sub-millimeter special resolution. Inspection of the SEM images of the sample surface in Fig. 8 shows small ellipsoidal voids where off-gassing of reduction products (H$_2$, CO) could occur through the SiC-containing layer. Micrographs indicate that gas-liquid (or gas-solid) reactions dominated over vapor phase-reactivity as no whisker formation was observed. The application of the volume-optimized composite precursor system improved surface morphology for SiC-AM fabrication as two characteristic microstructural defects of the single-component SLRS precursor materials were circumvented:

1. The mud-cracking of the surface produced by SiO$_2 \rightarrow$ SiC SLRS was no longer present. This may be attributed to compensatory volume changes for the Si/SiO$_2$ precursor.
2. The visibly distinct laser scan tracks associated with the formation Si$\rightarrow$ SiC SLRS were far less pronounced for conversion of Si/SiO$_2$ despite higher laser energy density. The addition of SiO$_2$ might allow laser light to penetrate the powder before being absorbed by Si. Through this process, laser energy may be more evenly distributed between adjacent scan tracks due to the melting pool broadening as particles are converted.

Cross-sectional SEM microscopy images shown in Fig. 10 indicate that the 4 W laser power heated the Si/SiO$_2$ very close to the melting point of Si (1410 °C). The lower melting point of Si compared to SiO$_2$ (1710 °C) favored partial or full melting of Si where it served as a liquid phase binder phase for converted SiO$_2$ particles within the irradiated layer. Minor power fluctuations during laser processing induced both solid-state/partial-melting behavior (see Fig. 10a) or solution-precipitation type conversion to SiC (see Fig. 10b). Full melting of the precursor was observed at the beginning of the raster scan path before laser energy could be attenuated by spontaneous carbon deposition or vaporized products (very minor compared to SiO SLRS) in the gas stream. The "adhesion depth" (particle bonding in the 3D structure) and the "SLRS-layer thicknesses" (2-dimensional continuity) were positively correlated with greater laser energy densities. Generally, these results indicate the potential for forming dense SiC structures using a full-melting, solution-precipitation mechanism. Overall, the SiC yields in this study were higher than those previously reported SLRS studies, and the surfaces that were produced had lower roughness [35].

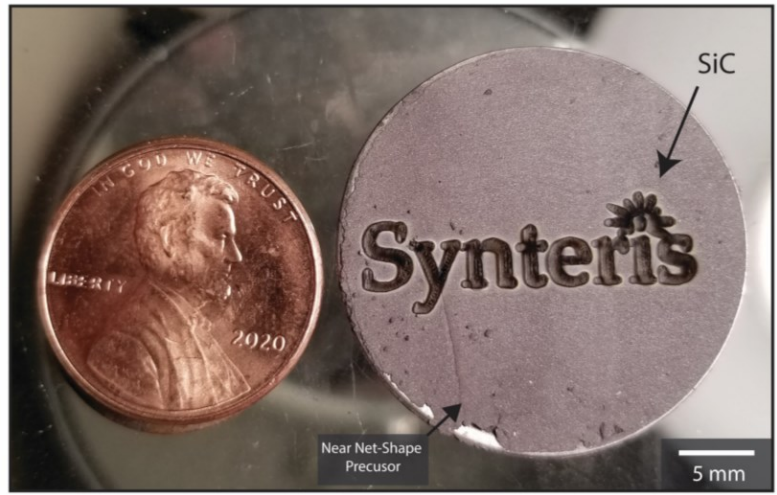

**Fig. 9**. Lettering selectively laser reaction sintered from near net-shape SiC precursors showing the sub-millimeter special resolution.

*2.2.3.2 Laser processing of near-net-shape Si/SiO in NH$_3$*

SLRS of the 69/31 wt% Si/SiO precursor material resulted in high Si$_3$N$_4$ yields of approximately 95%, however, surface and cross-sectional micrographs in Fig.7 and Fig. 10c indicate that Si$_3$N$_4$ layer continuity or dense nitride product was not achieved. The increase in laser power (from 3 W to 4 W) exacerbated materials vaporization and enhanced vapor-phase reactivity of SiO$_{(g)}$ with NH$_3$. Cross-sectional micrographs show that laser ablation of Si/SiO produced depressions in the layer surface and long Si$_3$N$_4$ whiskers. During processing, the reaction vessel visibly filled with opaque gaseous species from vaporization that attenuated laser power reaching the powder bed. As power output was increased from 4 W to 5.25 W, attenuation from SiO$_{(g)}$ interference was so significant that it reduced the Si$_3$N$_4$ yield and created microstructure inhomogeneity and spatially-varied whisker morphologies that are shown in Fig. 8c,d. Increased gas flow rates of 500 SCCM were not sufficient to quickly remove these materials from the reaction site.

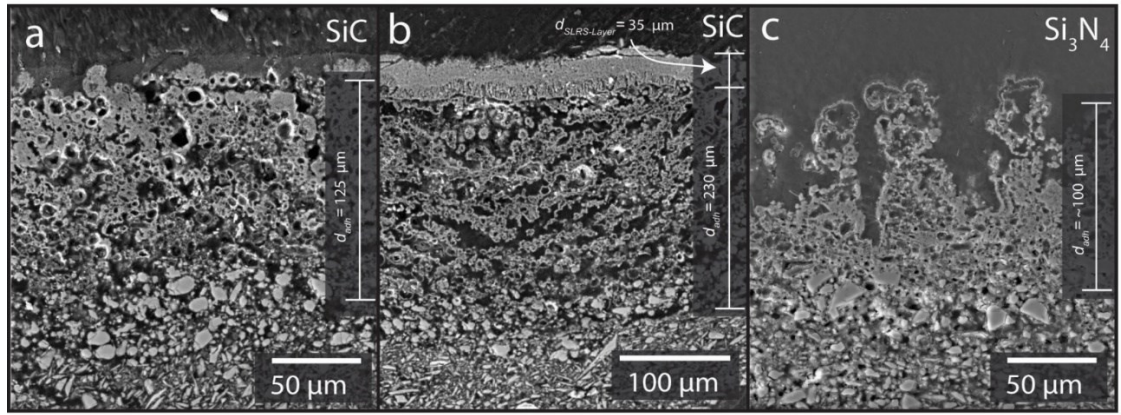

**Fig. 10.** SEM micrographs of epoxy encapsulated cross-sections of Si/SiO$_x$ precursor materials laser processed to Si$_3$N$_4$ (a) or SiC (b,c) by reaction with CH$_4$ or NH$_3$.

The SLRS α/β- Si$_3$N$_4$ yield presented in this study (~3:1 α/β- Si$_3$N$_4$ from Si or Si/SiO) was greater than that reported by Marcus and Birmingham (~1:2 α/β- Si$_3$N$_4$ of Si, 85 wt% yield); this ratio indicates reaction conditions favoring vapor-phase reaction over gas-liquid reactivity [43], [62], [63] The power density achieved with the 9 W, 1064 nm laser used by Marcus and Birmingham is unknown, however, it may be assumed that lower surface temperatures were achieved, favoring the lower α/β ratio and reducing non-oxide conversion (~85 wt% vs. ~95 wt%) [43]. For SLRS processing of Si$_3$N$_4$ using Si or SiO$_x$ precursors, there appears to be the following trade-off: high local temperatures increase nitride yield but also lead to whisker growth and unsatisfactory layer morphologies; low local temperatures prevent consolidation and result in an incomplete non-oxide conversion..

### 2.2.4 *Laser Synthesis of HfC/SiC Composite Materials*

Results in 2.2 suggest that metal and metal oxide species react through different mechanisms and rates. A composite system might be formed by reactive species by simultaneously converting metal and oxide species that have different base metal atoms. Results in 2.3.1.2 illustrated the high conversion efficiency of SiO$_2$ to SiC by SLRS processing in CH$_4$ without whisker formation. Similarly, our previous studies in [70] indicated approximately 80 wt% HfC$_{0.85}$ could be produced by SLRS processing of this same precursor material using 3 W laser power. The addition of a highly absorbing, refractory metal might aid conversion during SLRS synthesis. HfC-SiC composites are considered as potential candidates for thermal protection systems of hypersonic atmospheric re-entry vehicles [7]. The addition of additives and formation of HfC-SiC nanocomposites improve the sinterability, mechanical properties, and oxidation resistance of HfC UHTC materials [3], [7], [8]. In particular, the combination of high conversion efficiency of Hf to HfC and SiO$_2$ to SiC using SLRS processing and low vapor pressures (Fig. 6) makes them good candidates for HfC/SiC composite precursors for AM compatible reaction synthesis techniques.

The composite 94/6 wt% Hf/SiO$_2$ precursor material was laser processed in 100 vol% CH$_4$ using the methods and experimental apparatus described in 2.2.1. XRD results in Table 9 and Fig. 11c show high HfC/SiC composite yield (84.2 wt%), with 15.7 wt% Hf remaining unreacted. SiC was determined to account for ~5.8 wt% of the total crystalline mass. Only the cubic-3C polymorph of SiC was definitively detected. While not homogeneous on the microscale, the volume of SiC in the reacted Hf/SiC product materials is roughly 21.8 vol%, a value similar to the suggested range for oxidation-resistant refractory applications [3], [8]. In future iterations, the small additions of ternary HfO$_2$, or Si phases could be added to the precursor to tune the ratio of product non-oxides while preserving conversion-induced volume changes.

The surface microstructure of the HfC-SiC composite shown in Fig. 11a is nearly crack-free. Optical images and SEM micrographs show features that were not highly indicative of the conversion-induced volume changes previously reported for photothermochemical conversion of Hf to HfC [70]. Still, complete conversion of Hf to HfC through the optimization of SLRS processing parameters might further reduce residual stresses and improve layer morphology. The single-layer structure was robust enough to be removed from the powder bed, handled and cross-sectioned after epoxy encapsulation. SEM image analysis of the surface microstructure and layer-cross section indicate that particles adhered through solid-state reaction sintering processes rather than gas-liquid, or melt-precipitation type conversion for conversion of Si to SiC. The thickness of the reaction-bonded layer was determined to be ~35 μm which is within the typical range of melt depths for a laser powder bed fusion process (20-175 μm [71], [72]–[75]).

**Table 9. X-Ray Composition Analysis Using Rietveld Refinement Modeling: Conversion of Hf/SiO$_2$ Composite Precursor**

| Solid Phase Precursor (wt%) | Target Composite Product (Reactions Below) | Laser Power (W) | HfC (wt%) | SiC (wt%) | Unr. Hf (wt%) | Unr. SiO$_2$ (wt%) | Carbon Residue |
|---|---|---|---|---|---|---|---|
| Hf/SiO$_2$ (94/6 wt%) | HfC/SiC<br><br>Hf + CH$_4$ → **HfC** + 2H$_2$<br>+<br>SiO$_2$ + 3CH$_4$ → **SiC** + 2CO + 6H$_2$ | 5.25 | 78.4% | 5.8% (3C) | 15.8% | - | Yes |

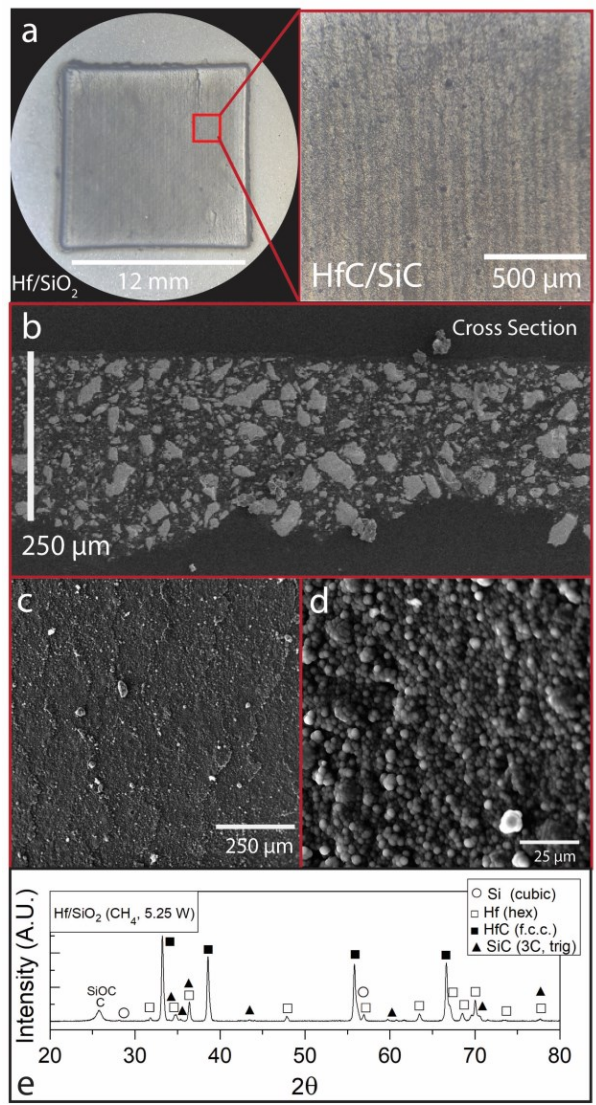

**Fig. 11.** (a) Photomicrographs and of near net-shape Hf/SiO$_2$ precursor laser sintered in 100 vol% CH$_4$ for HfC/SiC formation respectively. (b) SEM cross-section of the epoxy encapsulated cross-section. XRD spectrum of near net-shape Hf/SiO$_2$ processed using selective laser reaction sintering in 100 vol% CH$_4$. (c) and (d) surface morphology of HfC/SiC product layer. (e) x-ray diffraction spectra of product material.

While no-mechanical properties testing was performed, the high apparent density (that could be related to the broad nearly bimodal distribution of ~5 μm SiO$_2$ and ~20-40 μm Hf particles) might be promising for structural applications as samples with thicknesses of ~250 μm were able to be removed from their powder beds and handled. A combination of post-SLRS sintering at high temperatures could be used to enhance interparticle adhesion after *in-situ* synthesis and part shaping. Compared to other powder bed fusion methodologies for ceramics AM that commonly obtain relatively low densities (~40-90 vol% for $Al_2O_3$ [76], [77]]; ~60-90 vol% for post-processed SiC [78]) and still require binder phases, post-sintering, and/or infiltration, the single-step SLRS methodology might provide a viable alternative for dense components of highly refractory compositions. Alternatively, the application of higher laser energy densities might be used to drive melt-precipitation formation to further consolidate the SLRS microstructure and improve mechanical properties, so long as vapor-phase reactions and $SiO_2$ decomposition are avoided. Clearly further research is required to examine the translation of SLRS techniques to AM of SiC-containing parts, however the results show that composite feedstocks formulations containing SiC-precursors can be used to form unique refractory compositions that are traditionally incompatible with additive manufacturing.

## 3. Conclusions

The covalent non-oxides, silicon carbide and silicon nitride were synthesized using SLRS of Si, SiO, and $SiO_2$ precursor powders in 100 vol% $CH_4$ or $NH_3$ depending on the desired product. The composite material, HfC/SiC, was also SLRS processed using a $Hf/SiO_2$ mixed precursor system in a $CH_4$ atmosphere. X-ray phase characterization indicated that laser-assisted processing of Si to SiC, $SiO_2$ to SiC, and Si to $Si_3N_4$ could produce significant product yields. A combination of 3C, 96R, 6H-SiC phases (up to total 100 crystalline wt% SiC), and α/β- $Si_3N_4$ (up to a total of 97.7 crystalline wt% $Si_3N_4$) were produced. For the Hf/SiC composite, the total yield was determined to be 85.8 wt%. These non-oxide yields are greater than previously reported in other SLRS studies [35], [43]. Broadly, results indicate that AM-compatible non-oxide synthesis routes from $SiO_2$ to SiC and $Si_3N_4$ involve the key intermediate SiO [79]. In comparison to SLRS of Si to SiC, $SiO_2$ to SiC, and Si to $Si_3N_4$, the reduction of SiO or $SiO_2$ to $Si_3N_4$ was not viable due to requirements for high nitridation temperatures. When subject to direct laser heating, the high vapor pressure of $SiO_{(s)}$ caused material vaporization that attenuated laser energy deposition and/or induced whisker formation that hindered acceptable AM processing outcomes.

From this study, there the several key results that are significant to the application of SLRS reaction synthesis techniques to SiC, $Si_3N_4$ and HfC/SiC composite additive manufacturing:

1. <u>SiC</u>: SLRS of Si and $SiO_2$ in $CH_4$ indicates that SiC formation can occur through gas-solid, gas-liquid, or vapor-state reactions. Using the reaction conditions in this work, gas-solid and gas-liquid conversion mechanisms appeared to dominate due to the relatively high diffusion coefficient of C in $Si_{liq}$ ($2-5 \times 10^{-4}$ $cm^2$). Precipitation of SiC from the $Si/SiO_2$ melt formed continuous, crack-free, and dense layers of 93.7 wt% SiC that were approximately 35 μm thick.

2. <u>$Si_3N_4$</u>: SLRS formation of $Si_3N_4$ proved to be most efficient for nitridation of Si (97.7 wt%) in $NH_3$. Nitrogen has no solubility in solid Si and is slow to diffuse through liquid Si ($D_{N\ in\ Si\ (liq)}$: $3.2 \times 10^{-8}$ $cm^2/s$), α- $Si_3N_4$ ($D_{N\ in\ α-Si3N4\ (liq)}$: $1.2 \times 10^{-12}$ $cm^2/s$) and β- $Si_3N_4$ ($D_{N\ in\ β--Si3N4\ (liq)}$: $6.8 \times 10^{-6}$ to $1 \times 10^{-10}$ $cm^2/s$). The $Si_3N_4$ formation mechanism is then predicated on gas-liquid or vapor-state reactions. The high ratio of α/β- $Si_3N_4$ from Si and Si/SiO precursor materials indicates that conversion between Si and SiO and $NH_3$ occurred primarily in the vapor state. This gas-phase reaction did not produce the same conversion-induced volume changes seen in other systems.

3. <u>HfC/SiC composites:</u> SLRS formation of refractory composites from $Hf/SiO_2$ precursor in $CH_4$ showed high HfC/SiC composite yield (84.2 wt% total; 78.4 wt% HfC, 5.8 wt% SiC ), with 15.8 wt% Hf remaining unreacted. Broadly, the results indicate that SLRS processing can enable the formation of novel refractory composite compositions while still employing compensatory volume changes upon gas-solid (or gas-liquid reactivity).

## 4. Declaration of Competing Interest

Funding for the study described in this publication was provided by the Office of Naval Research. Under a license agreement between Synteris LLC and the Johns Hopkins University, Dr. Peters and the University are entitled to royalty distributions related to the technology described in the study discussed in this publication. Dr. Peters is a founder of and holds equity in Synteris LLC. He also serves as the Chief Technical Officer and holds a board seat to Synteris LLC. The results of the study discussed in this publication could affect the value of Synteris. This arrangement is pending review and approval by the Johns Hopkins University in accordance with its conflict of interest policies. Dennis Nagle, and Dajie Zhang are entitled to royalty distributions related to the technology described in the study discussed in this publication. Dr. Nagle, and Dr. Zhang are not affiliated with Synteris LLC. This arrangement is approved by the Johns Hopkins University in accordance with its conflict of interest policies.

## 5. Acknowledgments

The authors gratefully acknowledge funding provided by the Office of Naval Research, Nanomaterials Program Office, under contract N00014-16-1-2460 in partial support of this research. We would also like to thank the Johns Hopkins Applied Physics Laboratory Graduate Fellowship Committee for graduate student support provided for A.B. Peters.